\documentclass[reprint,superscriptaddress,aps,pra]{revtex4-2}
\usepackage{amsmath}
\usepackage{amssymb}
\usepackage{units}
\usepackage{graphicx,subfigure}
\usepackage{mathrsfs}
\usepackage{bm}
\usepackage{textcomp}
\usepackage{url}
\usepackage{dsfont}
\usepackage{braket}
\usepackage[compact]{titlesec}
\usepackage[colorlinks=true,linkcolor=magenta,citecolor=blue]{hyperref}
\usepackage{todonotes}
\usepackage{adjustbox}
\usepackage{empheq}
\usepackage{multirow}
\usepackage{ulem}
\usepackage{color}
\usepackage{lipsum}
\begin{document}
\title{Nonlocal Nonlinear Control of Photonic Spin Hall Effect in Strongly Interacting Rydberg Media}
\author{Wenzhang Liu}
\affiliation{Ministry of Education Key Laboratory for Nonequilibrium Synthesis and Modulation of Condensed Matter,
Shaanxi Province Key Laboratory of Quantum Information and Quantum Optoelectronic Devices, School of Physics, Xi’an Jiaotong University, Xi’an 710049, China}
\author{Muqaddar Abbas}
\email{muqaddarabbas@xjtu.edu.cn}
\affiliation{Ministry of Education Key Laboratory for Nonequilibrium Synthesis and Modulation of Condensed Matter,
Shaanxi Province Key Laboratory of Quantum Information and Quantum Optoelectronic Devices, School of Physics, Xi’an Jiaotong University, Xi’an 710049, China}
\author{Pei Zhang}
\email{zhangpei@mail.ustc.edu.cn}
\affiliation{Ministry of Education Key Laboratory for Nonequilibrium Synthesis and Modulation of Condensed Matter,
Shaanxi Province Key Laboratory of Quantum Information and Quantum Optoelectronic Devices, School of Physics, Xi’an Jiaotong University, Xi’an 710049, China}
\author{Jiawei Lai}
\email{laijiawei@xjtu.edu.cn}
\affiliation{Ministry of Education Key Laboratory for Nonequilibrium Synthesis and Modulation of Condensed Matter,
Shaanxi Province Key Laboratory of Quantum Information and Quantum Optoelectronic Devices, School of Physics, Xi’an Jiaotong University, Xi’an 710049, China}
\date{\today}
\begin{abstract}
We present a theoretical study demonstrating that the photonic spin Hall effect (PSHE) can be both actively tuned and greatly amplified by exploiting the nonlocal, nonlinear optical response of a strongly interacting Rydberg gas operating under electromagnetically induced transparency (EIT). Rather than relying on fixed phase metasurfaces or conventional Kerr materials, long range interactions between Rydberg atoms create a spatially extended, intensity dependent refractive index profile that enables dynamic spin resolved beam deflection. In a glass–Rydberg–glass trilayer model, we show that the PSHE displacement can be varied smoothly in magnitude and even reversed in sign by adjusting atomic density, probe intensity, laser detunings, and interaction strength. This nonlocal nonlinearity transforms the typically weak PSHE into a macroscopically observable and dynamically controllable effect. Our results establish a versatile platform for reconfigurable spin-dependent photonic control, with potential applications in high-sensitivity optical metrology, spin-based beam steering, and tunable quantum-optical information processing.
\end{abstract}
\maketitle
\section{Introduction}
\label{Sec:intro}

The photonic spin Hall effect (PSHE) is a spin-dependent lateral displacement that arises from the spin-orbit coupling of light beams at material interfaces or in inhomogeneous media \cite{bliokh2013goos}.  
Its magnitude in ordinary dielectrics is only a small fraction of the wavelength, so direct observation typically requires interferometric \cite{bliokh2008geometrodynamics} or weak-measurement techniques \cite{hosten2008observation,luo2011enhanced}.  
Nevertheless, the PSHE has garnered great interest as a sensitive light-matter interaction effect. 
For example, it has been used as an ultrafine optical metrology tool, by using the PSHE as a pointer in a weak measurement, researchers have achieved precision detection of monolayer graphene’s optical conductivity \cite{chen2020precision}.
To increase the PSHE shift, researchers have engineered metasurfaces and other nanostructures that impose tailored phase gradients \cite{ling2015giant,yu2024spatial}.  
A nanoantenna array, for example, has produced a giant PSHE by steering left- and right-circular polarizations in opposite directions \cite{yin2013photonic}. 
Such structures enable polarization-encoded information routing and precision metrology; spin-dependent beam shifts have been used to read nanometer surface features \cite{zahra2025spin} and to probe spin-orbit conversion in disordered media \cite{bardon2019spin}.  
However, one limitation of metasurface-based approaches is that the phase profile is fixed once fabricated.
Achieving an actively tunable PSHE, in which the spin-dependent splitting can be dynamically controlled, remains challenging.

Parallel to these developments, Rydberg atoms have emerged as an exciting platform for nonlinear and nonlocal optical physics.
Rydberg atoms are atoms excited to high principal quantum numbers, featuring enormous atomic orbitals and extremely strong long-range dipole-dipole interactions \cite{bai2016enhanced}.
A hallmark of Rydberg ensembles is the dipole blockade, in which one excited Rydberg atom can shift the energy levels of neighbors within a characteristic distance called blockade radius, and thereby simultaneous excitations are inhibited over micron-scale distances. 
In particular, under electromagnetically induced transparency (EIT) conditions a probe photon can be partially converted into a Rydberg polariton, acquiring a giant Kerr-type nonlinearity via Rydberg-Rydberg interactions \cite{sinclair2019observation}.
This spatially extended response modifies the refractive index tens of micrometers away, allowing contactless modulation of light propagation \cite{busche2017contactless}. 
At the few-photon level, Rydberg EIT has enabled single-photon switches and transistors, with F\"orster resonances boosting gain above 100, and two-photon gates achieving $\pi$ conditional phase shifts \cite{gorniaczyk2016enhancement,baur2014single}. 
Notably, increasing the degree of Rydberg interaction (e.g. by tuning atomic density or principal quantum number) can markedly change the spatial range of the nonlinear response, providing a knob to stabilize or manipulate nonlinear optical modes \cite{xu2020nonlocal}.
These advances establish Rydberg EIT as a unique platform for achieving and tuning nonlinear response that exceeds conventional materials by orders of magnitude, adjustable via atomic density, Rydberg level, or probe field intensity.

Building on these advances, in this paper we use an interacting Rydberg atomic medium to achieve an enhanced and tunable photonic spin Hall effect.
We consider a three-layer structure (glass-Rydberg-glass) in which a thin slab of Rydberg atomic gas under EIT serves as the central nonlinear layer. 
The nonlocal Kerr response produced by Rydberg-Rydberg interactions creates a spatially varying index perturbation that depends on the local intensity and polarization of the probe beam.  
This perturbation deflects photons of opposite spin by different amounts, amplifying the PSHE and critically making the shift a controllable parameter.  
Adjusting the probe laser power, the Rydberg level, or the atomic density changes the interaction strength and hence the spin displacement. 
The sign of the deflection can also be reversed by changing the detuning of the driving fields, a form of dynamic control unattainable with passive nanostructures. 

Despite extensive studies on the PSHE in various optical platforms, its behavior in planar thin-film trilayers hosting a strongly interacting Rydberg medium under EIT has remained unexplored.
In particular, the role of nonlocal, nonlinear optical responses arising from long-range Rydberg-Rydberg interactions on PSHE dynamics has yet to be addressed. This constitutes the central focus of the present work. Here, we investigate PSHE in a planar three-layer geometry, where a Rydberg atomic ensemble under ladder-type EIT serves as the nonlinear layer. Here, we investigate PSHE in a three-layer cavity geometry where a Rydberg atomic ensemble serves as the nonlinear medium under ladder-type EIT conditions. Our findings reveal that the PSHE shift can be continuously and precisely controlled in both magnitude and direction by tuning the probe-field detuning, with symmetric extrema emerging at positive and negative frequency offsets.

In contrast to earlier approaches that rely on fixed nanostructures or embedded atomic layers to enhance PSHE \cite{abbas2025tunable, waseem2024gain}, our scheme uniquely exploits interaction induced nonlocality in Rydberg gases to achieve dynamically tunable spin-dependent beam shifts, without the need for auxiliary gain media or complex multilevel configurations. This enables a robust, all-optical mechanism for real-time control of PSHE. 

In summary, our work introduces a distinct, interaction-mediated route to PSHE control that differs from prior approaches in three aspects. First, unlike metasurface or passive layered strategies that fix the phase profile after fabrication\cite{yin2013photonic,ling2015giant}, our mechanism exploits the nonlocal Kerr response generated by Rydberg–Rydberg interactions under EIT, enabling real-time, all-optical tuning of both the magnitude and sign of the PSHE near the Brewster condition. Second, in contrast to gain-assisted or specific multilevel interference schemes that remain essentially local in susceptibility, our control knobs—atomic density $N_a$, coupling Rabi frequency $\Omega_c$, and probe detuning $\Delta_2$—reshape the refractive index over the blockade radius, providing robust bidirectional control without auxiliary gain media\cite{abbas2025tunable,waseem2024gain}. Third, by explicitly incorporating the nonlocal term of $\chi^{(3)}$ into an angular-spectrum plus transfer-matrix framework for the glass–Rydberg–glass trilayer, we predict micrometer-scale shifts approaching $w_0/2$ and detuning-driven sign reversals at a fixed incident angle, thus offering an experimentally practical recipe for alignment-tolerant PSHE devices.

By integrating interacting Rydberg atoms into a planar thin-film trilayer, our work introduces a new route to enhance and manipulate the PSHE beyond the limitations of linear optics or conventional EIT media. The demonstrated tunability and amplification of spin-dependent light deflection not only advance fundamental understanding of light-matter interaction in the nonlocal nonlinear regime, but also pave the way for spin-resolved beam steering, precision optical metrology, and reconfigurable photonic information processing in next-generation quantum and spin-optical devices.

\begin{figure*}
    \centering
    \includegraphics[width=0.9\linewidth]{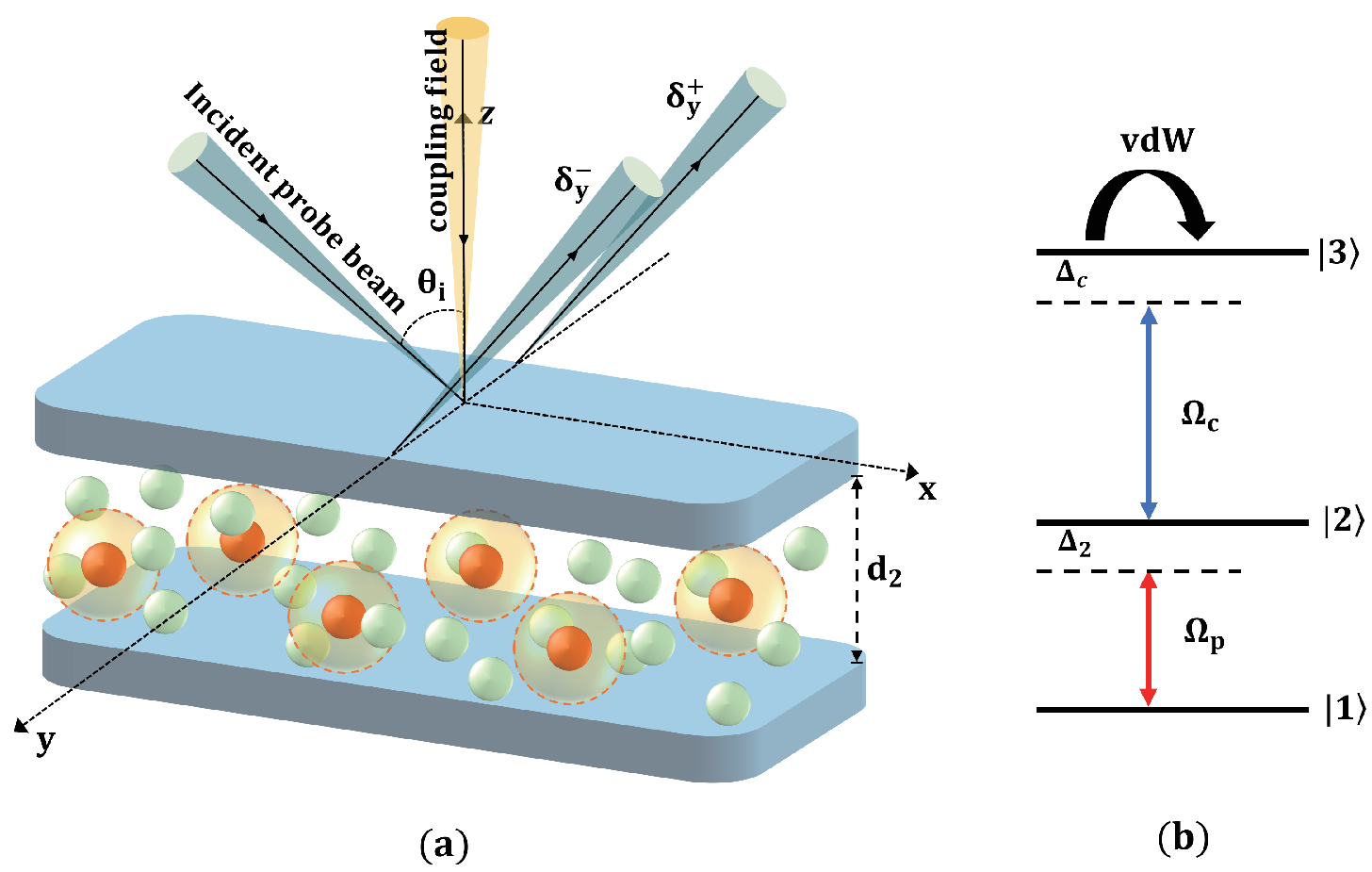}
    \caption{(a) Schematic of the planar three-layer (glass–Rydberg–glass) optical model. In experiment, the two “glass” layers correspond to the ultra-high-vacuum (UHV) cell windows, and the central layer is the laser-cooled $^{87}$Rb cloud (magneto-optical trap (MOT), optionally a weak dipole trap) inside the evacuated cell. The “layer thickness” $d_2$ denotes the axial extent of the probed atomic cloud inside the UHV cell. (b) Ladder-type EIT diagram defining the probe detuning $\Delta_{2}$ and two-photon detuning $\Delta_{3}=\Delta_{2}+\Delta_{c}$.}
    \label{fig:scheme}
\end{figure*}
\begin{figure*}
    \centering
    \includegraphics[width=0.9\linewidth]{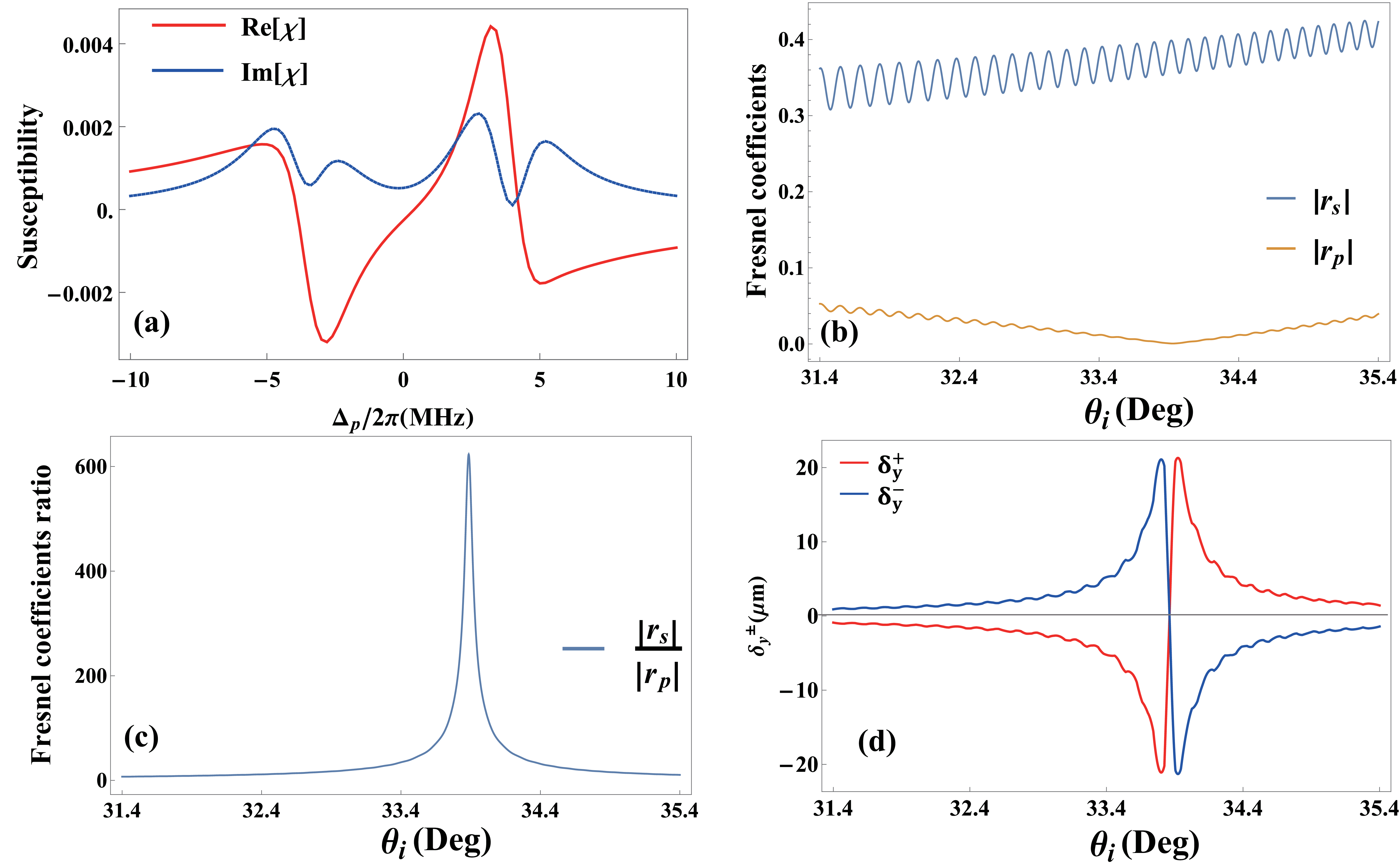}
    \caption{(a) The graphical representation of $\text{Re}[\chi]$ (red solid line) along with $\text{Im}[\chi]$ (blue dashed line) against the probe detuning $\Delta_2$. (b) Corresponding magnitudes of the Fresnel reflection coefficients for s- and p-polarizations, $|r_s|$ (blue solid line) and $|r_p|$ (orange solid line). (c) Ratio $|r_{s}|/|r_{p}|$ versus probe incident angle $\theta_i$. (d) The PSHE shifts $\delta_{y}^{\pm}$ versus probe incident angle $\theta_i$ for left(red solid line) and right(blue solid line) circular polarized reflected beams. Other parameters used in calculation: $\Omega_c/2\pi=4.0$MHz, $\Omega_p/2\pi=0.75$MHz, $N_a=4\times10^7$mm$^{-3}$, $\lambda_p=780$nm.}
    \label{fig:PSHE_shift}
\end{figure*}
\section{Theoretical Model}
\label{Sec:model}

The PSHE is the phenomenon in which the left and right circular polarization components of a light beam acquire opposite geometric phases upon reflection or refraction at an interface, leading to equal-magnitude, opposite-sign transverse shifts. This arises from the spin-orbit interaction of light, which converts spin-dependent phase gradients into nanometer-scale lateral displacements orthogonal to the plane of incidence. In Fig.~\ref{fig:scheme}(a) we consider a probe beam incident at an angle $\theta_{i}$ onto a planar three-layer stack. The multilayer consists of two parallel glass windows (the UHV cell windows) with susceptibility $\chi_1$ that sandwich a Rydberg atomic layer (the laser-cooled cloud) of thickness $d_2$ whose complex susceptibility is denoted by $\chi_2$. Here $d_2$ should be read as the axial extent of the probed cloud; it is set experimentally by the MOT size (and, if used, a weak dipole-trap waist). Inside the atomic layer, individual Rydberg excitations (red dots) are enveloped by dipole-blockade spheres (orange dashed circles), within which further excitations are inhibited; unexcited ground-state atoms are shown as green dots. These correlated Rydberg excitations gives rise to a strongly nonlocal nonlinear third order optical response superimposed on the usual linear susceptibility of the medium. Thus leads to a large tunability of the magnitude of PSHE in our system. Throughout this work, the strong coupling field is modeled as a large-waist plane wave normally incident on the stack, uniformly dressing the atomic layer and entering the theory via $\Omega_c$ and $\Delta_c$ in $\chi_2$.

In Fig.~\ref{fig:scheme}(a), the $z$ axis is normal to the interfaces of the layered structure and $x0z$ surface is the incident plane. We denote subscript $a=i,r$ for incident and reflected beams and superscript $H$ and $V$ to mark the horizontal and vertical polarizations to the incident plane. The incident beam is considered as a parallel polarized monochromatic Gaussian beam for its simplicity, the electric field amplitude of such a beam can be written as
\begin{equation}
    E^{H}_{i}(x_i,y_i)=E_{0} \exp(-\frac{x_{i}^{2}+y_{i}^{2}}{\omega^{2}_{0}}),
\end{equation}
where $E_0$ is the electric field amplitude of probe beam and $\omega_0$ is the beam waist. To model beam propagation and compute the PSHE quantitatively, we decompose the incident field $E^{H}_{i}(x_i,y_i)$ into its angular spectrum
\begin{equation}
    \tilde{E} ^{H}_{i}(k_{ix},k_{iy})=\int dx_{i} dy_{i} E^{H}_{i}(x_i,y_i)\exp[-i(k_{ix}x_i+k_{iy}y_i)],
\end{equation}
Each plane-wave component $\tilde{E} ^{H}_{i}(k_{ix},k_{iy})$ is then treated independently. From boundary condition $k_{rx}=-k_{ix}$ and $k_{ry}=k_{iy}$, the angular spectrum of reflected beam can be obtained as \cite{luo2011enhancing}
\begin{equation}
    \left[\begin{array}{c}
    \tilde{E}_r^H \\
    \tilde{E}_r^V
    \end{array}\right]=\left[\begin{array}{cc}
    r_p & \frac{k_{r y}\left(r_p+r_s\right) \cot \theta_i}{k_0} \\
    -\frac{k_{r y}\left(r_p+r_s\right) \cot \theta_i}{k_0} & r_s
    \end{array}\right]\left[\begin{array}{c}
    \tilde{E}_i^H \\
    \tilde{E}_i^V
    \end{array}\right],
    \label{eq:reflection}
\end{equation}
where $r_p$ and $r_s$ are the Fresnel reflection coefficients of the horizontal and vertical polarization components and $k_0=2 \pi/ \lambda_{p}$ is the central wave number of incident probe beam.

Next, we calculate the Fresnel coefficients via a standard transfer matrix method \cite{wu2010highly}. The transfer matrix of the $j$-th layer is represented by a $2 \times 2$ matrix
\begin{equation}
    M_j=\left(\begin{array}{cc}
    \cos \delta_j & -i \frac{\sin \delta_j}{p_j} \\
    -i p_j \sin \delta_j & \cos \delta_j
    \end{array}\right),
\end{equation}
Here $\delta_j=k_0 n_j d_j \cos \theta_j$ is the optical phase thickness of the $j$-th layer, and $p_j$ is dependent on the polarization of incident beam as $p_{j}^{p}=\frac{n_j}{cos[\theta_j]}$ and $p_{j}^{s}={n_j}{cos[\theta_j]}$. Overall, the total transfer matrix is $M=\prod_j M_j$. Denoting the incident side impedance $p_0=p_1$ and exit side $p_N=p_3$, we have the Fresnel coefficients with elements of transfer matrix $M$
\begin{equation}
    \begin{aligned}
    & r=\frac{\left(M_{11}+M_{12} p_3\right) p_1-\left(M_{21}+M_{22} p_3\right)}{\left(M_{11}+M_{12} p_3\right) p_1+\left(M_{21}+M_{22} p_3\right)}, \\
    & t=\frac{2 p_1}{\left(M_{11}+M_{12} p_3\right) p_1+\left(M_{21}+M_{22} p_3\right)} ,
    \end{aligned}
    \label{eq:fresnel}
\end{equation}
In Eq.~\eqref{eq:fresnel}, $r$ and $t$ denote the Fresnel reflection and transmission coefficients for the chosen polarization; adopting $p_j^p=n_j/\cos\theta_j$ yields $r = r_p$, $t= t_p$, whereas adopting $p_j^s=n_j\cos\theta_j$ yields $r= r_s$, $t= t_s$. Substitute Eq.~\eqref{eq:fresnel} into Eq.~\eqref{eq:reflection}, then write angular spectrum components in spin basis as $\tilde{E}_r^{ \pm}=\left(\tilde{E}_r^{H} \mp i \tilde{E}_r^{V}\right) / \sqrt{2}$. Here, we must remind that the Fresnel coefficients can be expanded as a polynomial of $k_{ix}$, but we can obtain a sufficiently good approximation by confining the Taylor series to the zeroth order \cite{luo2011enhancing}.

By taking an inverse Fourier transformation of $\tilde{E}_r^{ \pm}$, we can obtain the amplitude of reflected field. The transverse shift for the left$(\sigma=+1)$ and right$(\sigma=-1)$ circular polarized components of reflected beam can be calculated by
\begin{equation}
    \delta_y^\sigma=\frac{\iint y_r\left|E_r^\sigma\left(x_r, y_r\right)\right|^2 d x_r d y_r}{\iint\left|E_r^\sigma\left(x_r, y_r\right)\right|^2 d x_r d y_r},
\end{equation}

As we can clearly see that the PSHE shift $\delta_y^\sigma$ strongly depend on the incident angle $\theta_{i}$ and the susceptibility of each layer, especially the susceptibility of Rydberg atomic medium $\chi_{2}$ in our case. 

In a conventional three-level Rydberg atomic system, the probe susceptibility is $\chi=\chi^{(1)}+\chi_{\rm local}^{(3)}\left\lvert E_{p}\right\rvert ^2$, in which has two contributions: the linear term $\chi^{(1)}$ and the local Kerr term $\chi_{\rm local}^{(3)}$. The linear term can create the transparency window but respond weakly to other experimental knobs but the coupling field. The second term $\chi_{\rm local}^{(3)}\varpropto N_{a}$ only adds a nonlinear correction scales linearly with atomic density $N_a$ which only leads to limited modulation of the total susceptibility. By contrast, a Rydberg system with many-body interaction adds a nonlocal third-order term $\chi_{\rm nonlocal}^{(3)}$ that scales as the square of the atomic density and extends over the blockade radius. Because the blockade radius and interaction strength depend on the Rydberg energy level, probe and coupling Rabi frequency, and atomic density, the nonlocal nonlinearity is highly versatile and can be dynamically tuned in situ. This feature gives interacting Rydberg Medium a decisive advantage in tailoring refractive and absorptive properties, enabling giant and tunable PSHE shifts that are unattainable in conventional three-level schemes.

We consider a three-level ladder-type EIT scheme in the Rydberg medium (Fig.~\ref{fig:scheme}(b)), in which a weak probe of frequency $ \omega_p $ drives the $ |1\rangle\!\leftrightarrow\!|2\rangle $ transition at resonant frequency $ \omega_{21}=\omega_2-\omega_1 $, while a strong coupling field of frequency $ \omega_c $ addresses the $ |2\rangle\!\leftrightarrow\!|3\rangle $ transition at resonant frequency $ \omega_{32}=\omega_3-\omega_2 $. The single-photon detunings are defined as $\Delta_{2}=\omega_{p}-\omega_{21}$ for the probe on $|1\rangle\!\leftrightarrow\!|2\rangle$ and $\Delta_{c}=\omega_{c}-\omega_{32}$ for the coupling on $|2\rangle\!\leftrightarrow\!|3\rangle$; the two-photon detuning is $\Delta_{3}=\Delta_{2}+\Delta_{c}=\omega_{p}+\omega_{c}-(\omega_{3}-\omega_{1})$. Spontaneous emission decay from state $ |\beta\rangle $ to $ |\alpha\rangle $ occurs at rate $ \Gamma_{\beta\alpha} $, with the corresponding coherence decay rate $ \gamma_{\beta\alpha}=\Gamma_{\beta\alpha}/2\ $. The nonlocal van der Waals interaction between Rydberg atoms in state $|3\rangle$ is displayed as a arc-shaped arrow above state $|3\rangle$. The vdW interaction potential is $V(r_{ij})=-C_6 /r_{ij}^6$, as a function of $r_{ij}=| \mathbf{r_i}-\mathbf{r_j}   |$ to describe the interaction strength between two Rydberg atoms at $\mathbf{r_i}$ and$\mathbf{r_j}$.

Based on the above discussions, the total Hamiltonian for a three-level ladder-type Rydberg atomic system under the dipole and rotating-wave approximations with vdW interactions is given by $\hat{H} = N_a \int \mathrm{d}^3 \mathbf{r} \, \mathcal{H}(\mathbf{r})$. Here, $N_a$ is the atomic density, and $\mathcal{H}(\mathbf{r})$ is the Hamiltonian density \cite{bai2016enhanced}
\begin{equation}
    \begin{aligned}
    \hat{\mathcal{H}}(\mathbf{r}) / \hbar= & -\Delta_2 \hat{\sigma}_{22}(\mathbf{r})-\Delta_3 \hat{\sigma}_{33}(\mathbf{r}) \\
    & -\left(\Omega_p \hat{\sigma}_{12}(\mathbf{r})+\Omega_c \hat{\sigma}_{23}(\mathbf{r})+\text { H.c. }\right) \\
    & +N_a \int d^3 \mathbf{r}^{\prime} \hat{\sigma}_{33}\left(\mathbf{r}^{\prime}\right) V\left(\mathbf{r}^{\prime}-\mathbf{r}\right) \hat{\sigma}_{33}(\mathbf{r}).
    \end{aligned}
    \label{eq:Hamiltonian}
\end{equation}
Here, $\hat\sigma_{\alpha\beta}(\mathbf r)=\lvert\beta(\mathbf r)\rangle\langle\alpha(\mathbf r)\rvert$ denotes the transition operator for $\alpha\neq\beta$ (and the projection operator for $\alpha=\beta$); the Rabi frequencies of the probe and coupling fields are defined by $\Omega_p=\mathbf p_{21}\cdot\mathbf E_p/(\hbar)$ and $\Omega_c=\mathbf p_{32}\cdot\mathbf E_c/(\hbar)$, respectively, where $\mathbf p_{21}$ and $\mathbf p_{32}$ are the corresponding dipole matrix elements; and the final term represents the Rydberg-Rydberg interaction, described by the van der Waals potential $V(\mathbf r'-\mathbf r)=C_6/|\mathbf r'-\mathbf r|^6$ between atoms at $\mathbf r$ and $\mathbf r'$. The integral term accounts for the energy shift of an atom in state $\lvert3\rangle$ at position $\mathbf r$ induced by its nonlocal van der Waals interaction with an atom at position $\mathbf r'$.

Applying the Heisenberg equation of motion $i\hbar\partial_t\hat\sigma_{\alpha\beta}(\mathbf r)=[\hat\sigma_{\alpha\beta}(\mathbf r),\hat{\mathcal{H}}(\mathbf{r})]$, one obtains the following coupled equations for the one-body correlators:

\begin{subequations}\label{eq:bloch}
    \begin{align}
      & -i \tfrac{\partial}{\partial t} \rho_{11} = -i \Gamma_{12} \rho_{22}-\Omega_p \rho_{12} +\Omega_p^* \rho_{21}, \\ 
      & -i \tfrac{\partial}{\partial t} \rho_{22} =
      -i \Gamma_{23} \rho_{33}
      +i \Gamma_{12} \rho_{22}
      +\Omega_p \rho_{12}
      -\Omega_p^* \rho_{21} \notag \\
      & \qquad \qquad -\Omega_c \rho_{23}
      +\Omega_c^* \rho_{32}, \\ 
      & -i \tfrac{\partial}{\partial t} \rho_{33} = 
      +i \Gamma_{23} \rho_{33}
      +\Omega_c \rho_{23}
      -\Omega_c^* \rho_{32}, \\ 
      & -i \tfrac{\partial}{\partial t} \rho_{21} = 
      -d_{21}\rho_{21}
      -\Omega_p(\rho_{22}-\rho_{11})
      +\Omega_c^*\rho_{31}, \\ 
      & -i \tfrac{\partial}{\partial t}\rho_{31} =
      -d_{31}\rho_{31}
      -\Omega_p\rho_{32}
      +\Omega_c\rho_{21} \notag \\
      & \qquad \qquad -N_a\!\int d^3r'V(\mathbf{r}'-\mathbf{r})\rho \rho_{33,31}(\mathbf{r}',\mathbf{r}), \\ 
      & -i \tfrac{\partial}{\partial t} \rho_{32} = 
      -d_{32}\rho_{32}
      -\Omega_p^*\rho_{31}
      -\Omega_c(\rho_{33}-\rho_{22}) \notag \\
      & \qquad \qquad -N_a\!\int d^3r'V(\mathbf{r}'-\mathbf{r})\rho \rho_{33,32}(\mathbf{r}',\mathbf{r}).
    \end{align}
\end{subequations}
where $\rho_{\alpha\beta}(\mathbf r,t)=\langle\hat\sigma_{\alpha\beta}(\mathbf r,t)\rangle$ is the one-body density matrix element, $d_{\alpha \beta}=\Delta_\alpha-\Delta_\beta + i \gamma_{\alpha \beta}$($\Delta_1=0,\alpha \neq \beta$) and $\rho \rho_{\alpha \beta, \mu \nu }(\mathbf{r'},\mathbf{r})=\langle \hat{\sigma}_{\alpha \beta}(\mathbf{r^\prime}) \hat{\sigma}_{\mu \nu}(\mathbf{r})\rangle$ represents the quantum correlation contributed from Rydberg-Rydberg interaction.

To extract the probe susceptibility $\chi$ and its higher-order contributions, we first relate it to the atomic coherence using the expression $\chi=\frac{N_a|\mathbf{p}_{21}|^2}{\varepsilon_0\hbar}\frac{\rho_{21}}{\Omega_p}$. An explicit form for $\rho_{21}$ is found by solving the coupled Bloch equations in Eq.~\eqref{eq:bloch}(a)--(f). These equations involve two-body correlators like $\rho_{33,31}(\mathbf{r}',\mathbf{r})$, and their dynamics are given by the Heisenberg equation of motion

\begin{equation}\label{eq:two body bloch}
    \begin{aligned}
    & -i \tfrac{\partial}{\partial t} \rho \rho_{33,31}(\mathbf{r'},\mathbf{r})= \\
    & \quad\left(d_{31}+i \Gamma_{23}-V\left(\mathbf{r}^{\prime}-\mathbf{r}\right)\right) \rho \rho_{33,31}(\mathbf{r'},\mathbf{r}) \\
    & \quad+\Omega_c\left(\rho \rho_{23,31}(\mathbf{r'},\mathbf{r})+\rho \rho_{33,21}(\mathbf{r'},\mathbf{r})\right) \\
    & \quad-\Omega_c^*\rho \rho_{32,31}(\mathbf{r'},\mathbf{r}) -\Omega_p\rho \rho_{33,32}(\mathbf{r'},\mathbf{r}) \\
    & \quad-N_a \int d^3 \mathbf{r}^{\prime \prime}\left\langle\hat{\sigma}_{33}\left(\mathbf{r}^{\prime \prime}\right) \hat{\sigma}_{33}\left(\mathbf{r}^{\prime}\right) \hat{\sigma}_{31}(\mathbf{r})\right\rangle V\left(\mathbf{r}^{\prime \prime}-\mathbf{r}\right). \\
    \end{aligned}
\end{equation}   
It is therefore expected that Eq.~\eqref{eq:two body bloch} involves three-body correlators (e.g.\ $\left\langle\hat{\sigma}_{33}\left(\mathbf{r}^{\prime \prime}\right) \hat{\sigma}_{33}\left(\mathbf{r}^{\prime}\right) \hat{\sigma}_{31}(\mathbf{r})\right\rangle=\rho \rho \rho_{33,33,31}(\mathbf r'',\mathbf r',\mathbf r)$) as well as additional two-body terms such as $\rho \rho_{23,31}(\mathbf r',\mathbf r)$, leading to an infinite hierarchy of coupled equations for one-body, two-body, three-body and so on. One way to close the system of equations for the correlators is to expand the correlators in the powers of the probe Rabi frequency $\Omega_p$ \cite{sevinccli2011nonlocal}. This perturbation method has been detailed discussed in Ref.~\cite{bai2016enhanced, zhang2018fast}. We then apply their method by making the following expansions: $\rho_{\alpha 1}=  \sum_{j=0} \Omega_{p}^{2j+1} \rho_{\alpha 1}^{(2j+1)}, \rho_{\alpha \beta}= \sum_{j=1}\Omega_{p}^{2j} \rho_{\alpha \beta}^{(2j)}, \rho_{11}=1+ \sum_{j=1} \Omega_{p}^{2j} \rho_{11}^{(2j)}$, and for two-body correlators $\rho \rho_{\alpha \beta, \mu \nu}=\sum_{j=2} \Omega_{p}^{2j} \rho \rho_{\alpha \beta, \mu \nu}^{(2j)}, \rho \rho_{1 \beta, \mu \nu}=\sum_{j=1} \Omega_{p}^{2j+1} \rho \rho_{1 \beta, \mu \nu}^{(2j+1)}, \rho \rho_{1 1, \mu \nu}=\sum_{j=1} \Omega_{p}^{2j} \rho \rho_{1 1, \mu \nu}^{(2j)}$. With initial conditions $\rho_{11}^{(0)}=1, \rho_{22}^{(0)}= \rho_{33}^{(0)}=0$. Substitute these expansions into Eq.~\eqref{eq:bloch} and collect the terms of same exponential of $\Omega_p$. We can solve Eq.~\eqref{eq:bloch} in the steady state $\partial_t \rho_{\alpha \beta}=0$: 
\begin{subequations}\label{eq:one body expansion}
    \begin{align}
     & \rho_{21}^{(1)}=-\frac{d_{31}}{-|\Omega_c|^2+d_{21}d_{31}}, \\
     & \rho_{21}^{(3)}=-\frac{d_{31}(\rho_{22}^{(2)}-\rho_{11}^{(2)})-\Omega_c^{*} \rho_{32}^{(2)}}{|\Omega_c|^2-d_{21} d_{31}} \notag \\
     & \qquad +\frac{\Omega_c^* N_a}{|\Omega_c|^2-d_{21} d_{31}} \int d^3 \mathbf{r}^{\prime} V(\mathbf{r}^{\prime}-\mathbf{r}) \rho \rho^{(3)}_{33,31}(\mathbf{r}^{\prime},\mathbf{r}), 
    \end{align}
\end{subequations}
here, $\rho_{22}^{(2)}$, $\rho_{11}^{(2)}$ and $\rho_{32}^{(2)}$ are second order one-body correlators that are independent of Rydberg-Rydberg interactions and spatial position; in Eq.~\eqref{eq:one body expansion}(b), the first term on the right-hand side yields the local third-order susceptibility $\chi_{\rm local}^{(3)}$, whereas the second term arises from Rydberg-Rydberg interactions and hence corresponds to the nonlocal susceptibility $\chi_{\rm nonlocal}^{(3)}$. 
The blockade radius $R_b$ is defined by equating the van der Waals interaction strength $|C_6|/R_b^6$ to the EIT linewidth $\delta_{\rm EIT}=|\Omega_c|^2/\gamma_{12}$ \cite{pritchard2010cooperative}, since that no second atom within $|\mathbf r'-\mathbf r|<R_b$ can be excited to the Rydberg state $|3\rangle$ due to the large interaction-induced level shift; accordingly, the spatial integral in the nonlocal term of Eq.~\eqref{eq:one body expansion}(b) extends from $|\mathbf r'-\mathbf r|=R_b$ to $\infty$, though in practice one may truncate the upper limit at $\sim 3 R_b$ because $V(\mathbf r'-\mathbf r) \varpropto  |\mathbf{r}'-\mathbf{r}|^{-6}$, which decays rapidly beyond the blockade radius \cite{zheng2022correlation}.

In Appendix~\ref{Sec:Appendix} we present the detailed perturbative derivation for the two-body correlator $\rho \rho_{33,31}^{(3)}$.

We choose a laser-cooled $^{87}$Rb atomic gas with atomic states $\lvert1\rangle\equiv\lvert5S_{1/2},F=2,m_F=2\rangle$, $\lvert2\rangle\equiv\lvert5P_{3/2},F=3,m_F=3\rangle$, $\lvert3\rangle\equiv\lvert60S_{1/2}\rangle$, with spontaneous decay rates $\Gamma_{21}=2\pi\times6$MHz, $\Gamma_{32}=2\pi\times3$kHz, and the corresponding van der Waals coefficient $C_6=2\pi\times140$GHz$\cdot\mu$m$^6$ \cite{pritchard2010cooperative}.
\begin{figure*}
    \centering
    \includegraphics[width=0.9\linewidth]{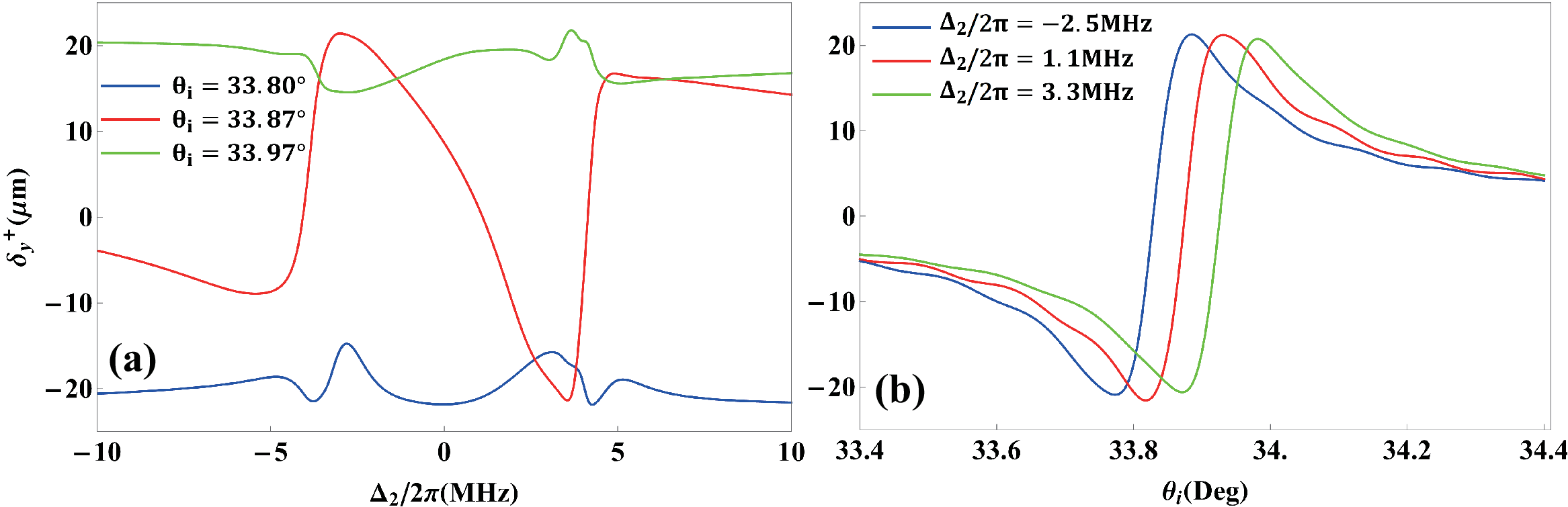}
    \caption{The PSHE shift for left circular polarized beam $\delta_y^{+}$.(a)Shift versus probe detuning $\Delta_{2}/2\pi$ for three incident angles around the Brewster value $\theta_{i}=33.80^\circ$(blue), $\theta_{i}=33.87^\circ$(red) and $\theta_{i}=33.97^\circ$(green). (b)Shift versus incident angle $\theta_i$ for three representative detunings $\Delta_{2}/2\pi=-2.5$MHz(blue), $\Delta_{2}/2\pi=1.1$MHz(red) and $\Delta_{2}/2\pi=3.3$MHz(green).   Other parameters used in calculation: $\Omega_c/2\pi=4.0$MHz, $\Omega_p/2\pi=0.75$MHz, $N_a=4\times10^7$mm$^{-3}$, $\lambda_p=780$nm.}
    \label{fig:PSHE_shift_detuning_angle}
\end{figure*}
\section{Results}\label{Sec:results}

In this section, we investigate the PSHE under the condition that the coupling field is detuned by $\Delta_c/2\pi=-0.1$MHz while the probe detuning $\Delta_2$ is varied; this configuration preserves EIT transparency at resonance and suppresses undue probe absorption. We use experimentally realistic parameters chosen to match existing Rydberg EIT experiments, and we analyze the spin dependent transverse shifts using the angular spectrum and transfer matrix methods to clarify the mechanisms behind the enhanced tunable PSHE in our proposed system. We also compare the medium susceptibility with and without Rydberg-Rydberg interactions and examine the resulting PSHE under both scenarios, in order to clearly demonstrate the impact of our scheme on enhancing and controlling the PSHE.

To study PSHE modulated by a Rydberg atomic medium, we first investigate the optical response of the medium. Fig.~\ref{fig:PSHE_shift}(a) contrasts the susceptibility of the probe transition with Rydberg-Rydberg interactions as a function of the probe detuning $\Delta_2$.  
Fig.~\ref{fig:PSHE_shift}(a) shows that the entire dispersion slightly shifts towards higher frequencies and steepens both the negative and the positive peak, signalling an blue-shift of the Rydberg level and an enhanced group-index slope.  
In the imaginary part, Fig.~\ref{fig:PSHE_shift}(a) shows the standard Autler-Townes doublet splits into multiple asymmetric peaks, the central transparency window narrows and its floor rises, evidencing residual absorption.  
This susceptibility induced by interactions between Rydberg atoms can be tuned systematically by adjusting experimental parameters, such as $\Omega_p$ and the atomic density $N_a$. This tunability enables precise engineering of the medium's refractive index, which in turn provides an effective handle for modulating the photonic spin Hall effect of the reflected light.
Next, we examine the PSHE shifts $\delta_{y}^{\pm }$ as a function of incident angle $\theta_i$ around the Brewster angle $\theta_{B}\approx \text{arctan}(n_2/n_1) \approx 33.8^\circ$(glass refractive index $n_1=1.49$) at detuning $\Delta_2=0$. 
By setting the thickness of atomic medium $d_2=100 \mu m$ and the incident beam waist $w_{0}=50 \mu m$, Fig.~\ref{fig:PSHE_shift}(d) shows that when away from $\theta_B$, both the lateral shifts of left(red) and right(blue) circular polarized components remain sub-wavelength level. 
Near $\theta_B$, both curves become sharp with enhanced PSHE shifts. The opposite peaks reach $|\delta_{y}^{\pm}|\approx 20~\mu\mathrm{m}$, close to the theoretical maximum of $w_0/2$.
The fine saw-tooth oscillations superimposed on these peaks arise from the interference within the three layer structure.
Fig.~\ref{fig:PSHE_shift}(b) shows the p- and s-polarized Fresnel reflection coefficients, $ |r_p(\theta_i)| $ and $ |r_s(\theta_i)| $, across the same angular range. As $\theta_i$ grows, $ |r_p| $ drops to zero at $\theta_B$ and then climbs again, whereas $ |r_s| $ rises monotonically.
Superimposed on these global trends are small-amplitude, high-frequency oscillations produced by multiple internal reflections within the three-layer structure. 
The distinct angular responses of $|r_p|$ and $|r_s|$ highlight the pronounced sensitivity of the light-interface interaction to the angle of incidence.

Fig.~\ref{fig:PSHE_shift}(c) plots the ratio $|r_s|/|r_p|$ as a function of $\theta_i$. 
This ratio diagnoses the imbalance between the $s$- and $p$-polarized reflection channels and thus serves as a proxy for the amplification of the PSHE near the Brewster region. 
From the reflection matrix in Eq.~\ref{eq:reflection}, the $H\!\leftrightarrow\!V$ spin-mixing element scales as $(r_p+r_s)\,k_{ry}\cot\theta_i/k_0$, whereas for a $p$-polarized input the background scales with $r_p$; hence the relative mixing strength increases roughly as $(r_p+r_s)/r_p$, well captured by the simpler indicator $|r_s|/|r_p|$. 
As $\theta_i\!\to\!\theta_B$, $|r_p|\!\to\!0$ while $|r_s|$ remains finite, producing a sharp peak in $|r_s|/|r_p|$ that marks the narrow angular window where the spin-dependent centroid shifts $\delta_y^{\pm}$ in Fig.~\ref{fig:PSHE_shift}(d) are maximized.

\begin{figure*}
    \centering
    \includegraphics[width=0.8\linewidth]{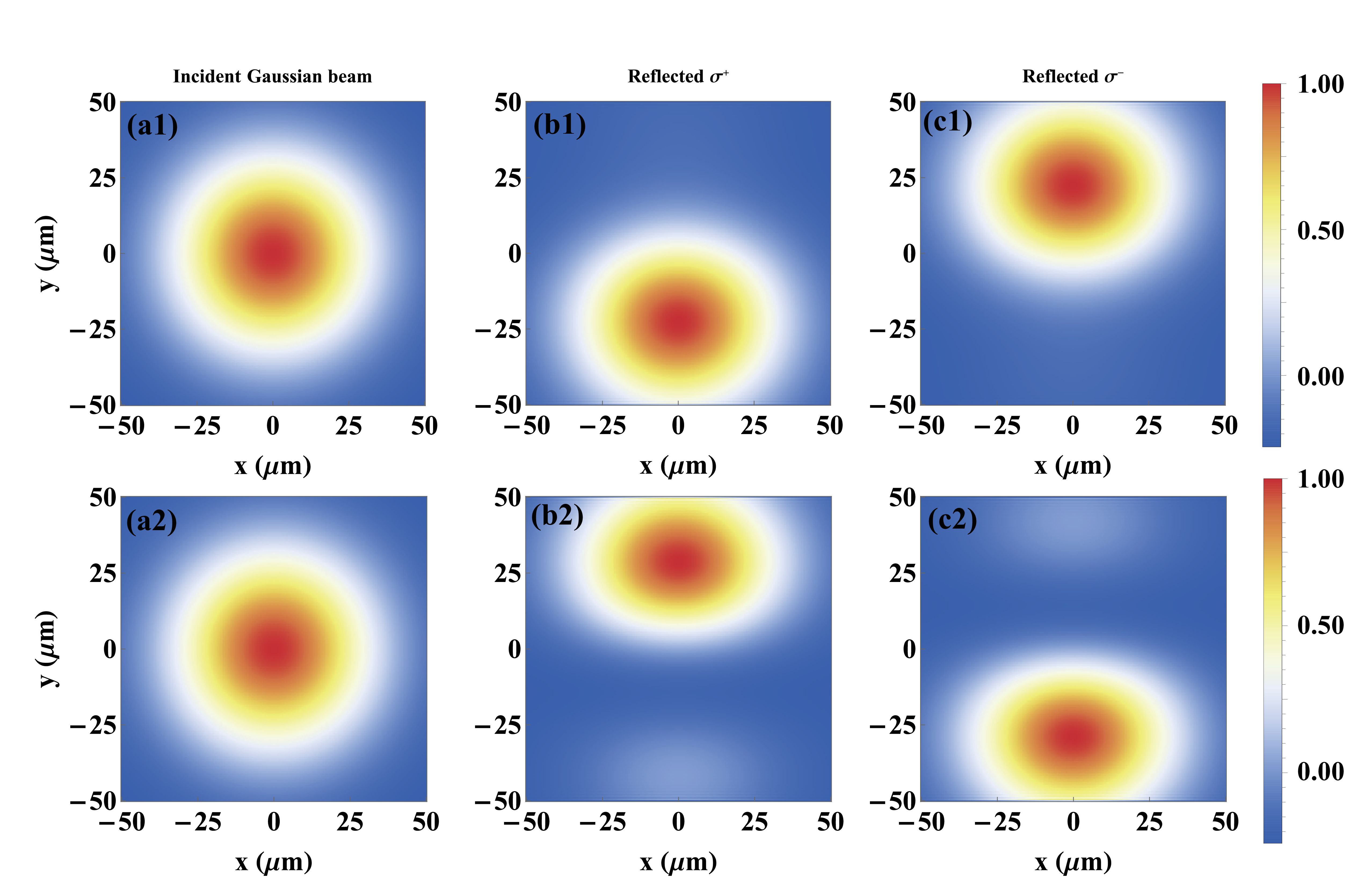}
    \caption{The distribution of normalized transverse intensity for (a)incident Gaussian beam (b) reflected left-circular polarized field and (c) reflected right-circular polarized field. The first row corresponds to probe detuning $\Delta_2 / 2 \pi= 3.5 $MHz and the second row corresponds to probe detuning $\Delta_2 / 2 \pi= -3 $MHz. Other parameters are the same in Fig.~\ref{fig:PSHE_shift}.}
    \label{fig:field-intensity-distribution}
\end{figure*}

\begin{figure*}
    \centering
    \includegraphics[width=0.8\linewidth]{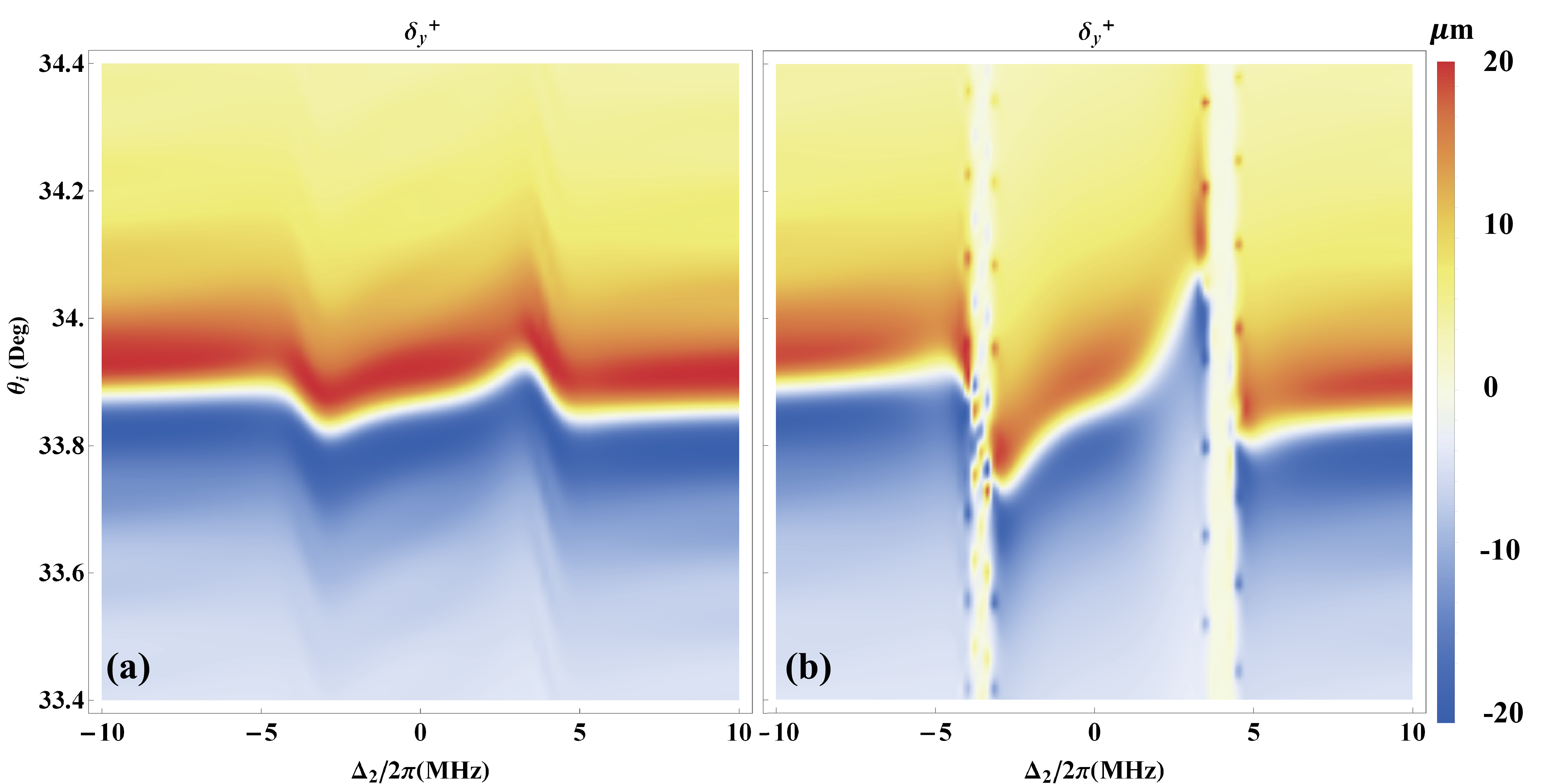}
    \caption{Density plot of PSHE shift $\delta_{y}^+$ versus incident angle $\theta_i$ and probe detuning $\Delta_2$ at different atomic density (a)$N_a=4 \times 10^{7}\text{mm}^{-3}$, (b)$N_a=8 \times 10^{7}\text{mm}^{-3}$. Other parameters are the same in Fig.~\ref{fig:PSHE_shift}.}
    \label{fig:density_PSHE}
\end{figure*}

\begin{figure*}
    \centering
    \includegraphics[width=0.7\linewidth]{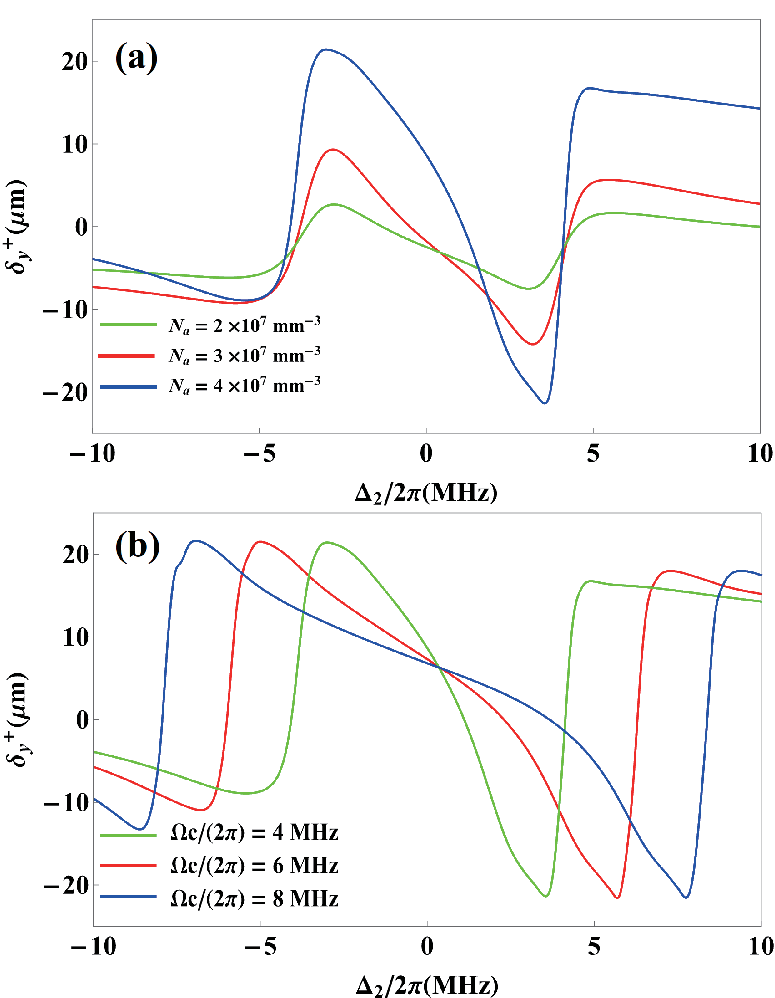}
    \caption{(a) The PSHE shift for left circular polarized beam $\delta_y^{+}$ versus probe detuning $\Delta_{2}/2\pi$ for three atomic densities  $N_a=2 \times 10^{7} \text{mm}^{-3}$(green), $N_a=3 \times 10^{7} \text{mm}^{-3}$(red) and $N_a=4 \times 10^{7} \text{mm}^{-3}$(blue). (b) The PSHE shift for left circular polarized beam $\delta_y^{+}$ versus probe detuning $\Delta_{2}/2\pi$ for  $\Omega_c/2 \pi=4 \text{MHz}$(green), $\Omega_c/2 \pi=6 \text{MHz}$(red) and $\Omega_c/2 \pi=8 \text{MHz}$(blue). Other parameters are the same in Fig.~\ref{fig:PSHE_shift}.}
    \label{fig:change_Na_PSHE}
\end{figure*}

To resolve the effects in the vicinity of the Brewster angle with high precision, we restrict the analysis to a narrow range of incident angles~$\theta_i$. Figure~\ref{fig:PSHE_shift_detuning_angle} shows how the Rydberg interaction modulates the PSHE shift: panel (a) plots the shift versus detuning at a fixed angle, while panel (b) shows it versus angle at a fixed detuning.
Note that in our glass–Rydberg–glass trilayer the Brewster angle $\theta_B$ is not fixed. 
It varies with the Rydberg-layer susceptibility $\chi_{2}$, and thus with atomic density and laser detunings.
For clarity, our discussion is confined to the transverse shift of the left circular polarized component, denoted $\delta_{y}^{+}$. 
Fig.~\ref{fig:PSHE_shift_detuning_angle}(a) shows $\delta_y^{+}$ as a function of probe detuning $\Delta_{2}/2\pi$ for three incident angles situated on either side of the Brewster angle $\theta_B \approx 33.8^{\circ}$.  
For $\theta_i = 33.80^{\circ}$ (blue), slightly below $\theta_B$, the shift remains negative, varying only between $-15$ and $-22\mu\text{m}$.  
At $\theta_i = 33.87^{\circ}$ (red), almost exactly at $\theta_B$, the detuning sweeps the shift through a pronounced sign reversal: it first reaches a positive maximum of $20\mu\text{m}$ near $\Delta_{2}/2\pi \approx -3\text{MHz}$ and then surges to a negative extremum of $-22\mu\text{m}$ around $+3\text{MHz}$, yielding a tunability window exceeding $40\mu\text{m}$ and in both directions approaching the theoretical limit $w_{0}/2$.  
When the angle is increased to $\theta_i = 33.97^{\circ}$ (green), the shift stays positive over the entire detuning range, fluctuating modestly about $+20\mu\text{m}$.
Hence, by selecting an incident angle close to the Brewster condition, one can exploit the probe detuning to tune the PSHE shift over a remarkably wide range of nearly $40\mu\text{m}$.
Fig.~\ref{fig:PSHE_shift_detuning_angle}(b) plots $\delta_y^{+}$ versus the incident angle $\theta_i$ for three probe detunings: $\Delta_{2}/2\pi = -2.5$MHz (blue), $+1.1$MHz (red), and $+3.3$MHz (green).  
For all curves the shift is initially negative, reaching minima of approximately $-20\mu$m, at $\theta_i \approx 33.8^{\circ}$.  
As $\theta_i$ increases past the Brewster angle $\theta_B\approx33.8^\circ$, a sign reversal occurs, each curve undergoes a sharp sign reversal and attains a positive maximum near $+20\mu$m.
Importantly, the angle at which this reversal occurs is detuning dependent, the peak moves from $\theta_i \approx 33.85^{\circ}$ for red detuning to $\approx 33.93^{\circ}$ for the blue detuning, corresponding to an angular shift of about $0.08^{\circ}$.  
This frequency controlled displacement of the PSHE extremum provides a practical means of compensating angular misalignment in experiments, allowing the spin-dependent beam separation to be optimised without mechanical readjustment.
Overall, Fig.~\ref{fig:PSHE_shift_detuning_angle} demonstrates that simultaneous tuning of the incident angle and the probe detuning provides bidirectional and large-amplitude control of the PSHE shift, with maximum sensitivity attained in the immediate vicinity of the Brewster angle.

Fig.~\ref{fig:field-intensity-distribution} complements Fig.~\ref{fig:PSHE_shift_detuning_angle}(a) by mapping the normalized transverse intensity profiles of the reflected beam at two representative detunings. At $\Delta_2 / 2 \pi = 3.5$MHz the left-circular component peaks at $y = -20 \mu\text{m}$ while the right-circular component peaks at $y = +20 \mu\text{m}$; changing the detuning to $-3 $MHz reverses these positions, directly visualizing the detuning controlled sign flip of the PSHE shift.

In Fig.~\ref{fig:density_PSHE} we present density plots of the PSHE shift as functions of the incident angle $\theta_i$ and the probe detuning $\Delta_{2}$ with different atomic density.  
Fig.~\ref{fig:density_PSHE}(a) gives the left circular polarized component $\delta_{y}^{+}$ with $N_a=4 \times10^{7} \text{mm}^{-3}$, whereas Fig.~\ref{fig:density_PSHE}(b) of $N_a=8 \times10^{7} \text{mm}^{-3}$. 
A narrow zero-shift ridge divides the positive and negative displacement domains.  
Its centre lies at $\theta_i\simeq33.8^{\circ}$ for zero detuning and is tilted such that red detuning shifts the ridge toward smaller angles, whereas blue detuning shifts it toward larger ones.  
The tilt amounts to approximately $0.08^{\circ}$ over a detuning span of $\pm3$MHz, enabling frequency-controlled steering of the Brewster-like working angle.  
On either side of this ridge, the displacement attains extrema of about $\pm20 \mu$m.  
The extrema appear within an angular bandwidth of $\sim0.2^{\circ}$, defining the tolerance window for maximizing the spin-dependent splitting.  
Fig.~\ref{fig:density_PSHE}(b) breaks the detuning dependence of PSHE, shows that the enhanced nonlinearity sharpens the zero-shift ridge and expands the tunable angle window. However it also fragments the formerly smooth detuning response, so the shift can no longer be tuned continuously with $\Delta_{2}$.
These plots underline the enhancement and tunability of the PSHE. 
Adjusting $\Delta_2$ can compensate mechanical angle errors. 
Conversely, tuning $\theta_i$ can offset laser-frequency drifts. 
Together, these knobs provide a robust route to maintain optimal spin separation.

Finally, we emphasise that tunability of the PSHE shift relies critically on Rydberg-Rydberg interaction.
Fig.~\ref{fig:change_Na_PSHE}(a) plots the PSHE shift $\delta_y^{+}$ as a function of the probe detuning $\Delta_{2}$ for three atomic densities $N_{a}=2\times10^{7}$, $3\times10^{7}$, and $4\times10^{7}\,\mathrm{mm^{-3}}$; at the lowest density (green) the shift varies only weakly, remaining between $-10$ and $+3\,\mu\mathrm{m}$; raising the density to $3\times10^{7}\,\mathrm{mm^{-3}}$ (red) broadens the response and produces a sign reversal with a tunability window of nearly $20\,\mu\mathrm{m}$; at $4\times10^{7}\,\mathrm{mm^{-3}}$ (blue) the shift exceeds $40\,\mu\mathrm{m}$, reaching extrema of $-20$ and $+22\,\mu\mathrm{m}$ that approach the theoretical limit $w_{0}/2$. 
The strong density dependence originates from the $N_{a}^{2}$ scaling of the nonlocal third-order susceptibility\cite{sevinccli2011nonlocal}. As the atomic density increases, the Rydberg-Rydberg interaction enhances $\chi_{\mathrm{nonlocal}}^{(3)} \propto N_{a}^{2}$. This results in a rapid, interaction-induced change of the refractive index, which strongly modulates the PSHE shift.
When either the atomic number density $N_a$ is small or the probe Rabi frequency is weak, the interaction-induced nonlocal third-order contribution becomes negligible so that the overall response effectively reduces to conventional three-level EIT; correspondingly, the lower-$N_a$ curve in Fig.~\ref{fig:change_Na_PSHE}(a) exhibits only limited PSHE tunability under parameter changes. 
In contrast, for larger $N_a$ (blue curve), Rydberg blockade enhances the nonlocal contribution and reshapes $\mathrm{Re}[\chi]$, enabling both amplified modulation and robust sign reversals at the same fixed angle.
In Fig.~\ref{fig:change_Na_PSHE}(b) we demonstrate the impact of other experimental parameter such as coupling field Rabi frequency, as we increase $\Omega_c$, starting with $\Omega_c/ 2 \pi =4.0 \text{MHz}$ to $\Omega_c/ 2 \pi =8.0 \text{MHz}$, it shows a broadened tuning windows of probe detuning $\Delta_2$, of which can be applied in experiments to compensate the accuracy of detuning.

Building on these results, interacting Rydberg atomic medium offers clear benefits in real-time PSHE shift adjusting. 
The nonlocal Kerr term scales with the square of the atomic density and extends across the blockade radius, yielding a nonlinear susceptibility that can exceed the linear term by orders of magnitude even at moderate densities and thereby enhancing PSHE shifts without high optical power. 
Because the interaction strength and susceptibility profile all depend on the chosen Rydberg state, the atomic density and the probe and coupling Rabi frequencies, tuning these parameters enables real-time control of the PSHE. 
These enhancements persist over wide detuning and incidence-angle ranges, providing a robust platform for ultrahigh-sensitivity displacement and refractive-index sensing that passive medium cannot match.

\section{Possible Experimental Realization of Proposed Model}\label{Sec:experimental realization}

A direct realization employs a standard ultra-high-vacuum glass cell hosting a cold $^{87}\mathrm{Rb}$ ensemble under ladder-type Rydberg EIT, where the two windows serve as the “glass” layers of our model and the laser-cooled cloud in between constitutes the Rydberg medium. The probe at $\lambda_{p}=780~\mathrm{nm}$ drives $|5S_{1/2}\rangle\!\leftrightarrow\!|5P_{3/2}\rangle$ with $\Omega_{p}/2\pi\sim0.5{-}1~\mathrm{MHz}$, while the coupling at $\lambda_{c}\approx480~\mathrm{nm}$ addresses $|5P_{3/2}\rangle\!\leftrightarrow\!|nS_{1/2}\rangle$ ($n\!\approx\!60$) with $\Omega_{c}/2\pi\sim4{-}8~\mathrm{MHz}$. Number densities $N_{a}\sim\mathrm{few}\times10^{7}\,\mathrm{mm^{-3}}$, axial thickness $d_{2}\sim100~\mu\mathrm{m}$, and probe waist $w_{0}\sim50~\mu\mathrm{m}$ are routinely achieved in cold-atom Rydberg-EIT experiments and match the operating points used in our figures. In the reflection geometry, the probe impinges on a window at an incidence near the trilayer Brewster angle $\theta_{B}$ (typical value for our glass parameters $\sim 33.8^\circ$), exactly as modeled in Sec.~\ref{Sec:model}. The coupling beam traverses the cloud with a larger waist to set $\chi_{2}$, but does not participate in the probe’s reflection path.

Detection uses a QWP$\to$PBS analyzer to separate the reflected $\sigma^{\pm}$ components\cite{Xiang2017PRJ}, which are then imaged on a low noise camera. We extract beam centroids along the transverse direction and obtain the differential spin Hall shift $\delta_{y}^{\pm}$. For the above parameters, our calculations (Figs.~\ref{fig:PSHE_shift}–\ref{fig:change_Na_PSHE}) predict spin dependent displacements up to $\pm(15{-}22)~\mu\mathrm{m}$ near $\theta_{B}$. These shifts are resolvable with standard pixel scales after modest averaging. Weak measurement amplification is optional\cite{Choi2024Nanophotonics,Zhou2012PRA}. We first calibrate the Brewster angle without atoms, where $r_{p}$ vanishes. We then fine tune the angle with the cloud present to account for the $\chi_{2}$ and $d_{2}$ dependence of $\theta_{B}$. All experimental ingredients are standard and have been demonstrated repeatedly, including laser wavelengths, detunings, Rabi frequencies, atomic densities, cloud dimensions, and the reflection based readout. This provides a direct and practical route to realize the tunable PSHE reported here.

\section{Conclusions}\label{Sec:conclusions}
In summary, we have presented a comprehensive theoretical study of the PSHE in a glass–Rydberg–glass sandwich cavity operating under a ladder-type EIT scheme. In summary, we have presented a comprehensive theoretical study of the PSHE in a glass–Rydberg–glass planar trilayer stack operating under a ladder-type EIT scheme. By expanding the atomic Bloch equations up to third order in the probe field, we incorporated both the conventional Kerr nonlinearity and the interaction-induced nonlocal nonlinear response arising from strong Rydberg-Rydberg interactions. This was integrated into a multilayer transfer matrix framework to capture spin-resolved beam dynamics.

Our analysis reveals that the nonlocal nonlinearity significantly modifies the optical susceptibility of the Rydberg medium, enabling large, tunable spin-dependent transverse shifts well beyond the limits of conventional PSHE platforms. Crucially, the interplay between detuning and interaction strength allows for the control of both the magnitude and sign of the PSHE shift at a fixed incident angle near the Brewster condition.

These results establish a new pathway for implementing reconfigurable, spin-sensitive photonic components. By leveraging frequency-controlled tuning rather than fixed nanostructures or complex geometries, the proposed scheme offers a versatile and scalable approach for applications in optical metrology, beam steering, and quantum-enhanced photonic signal processing. 
Compared to earlier PSHE platforms based on fixed nanostructures or local-Kerr media, our interaction-induced, nonlocal Rydberg mechanism enables macroscopically large and sign-reversible shifts at a fixed angle with in-situ reconfigurability~\cite{luo2011enhancing,yin2013photonic,ling2015giant}.

\appendix
\section{APPENDIX: DERIVATION OF PERTURBATION METHOD}\label{Sec:Appendix}
In addition to Eq.~\eqref{eq:two body bloch}, other dynamic equations of two-body correlators are
\begin{widetext}
\begin{subequations}\label{eq:two body bloch 1}
    \begin{align}
    & \left(i \frac{\partial}{\partial t} + d_{23}+ d_{31}\right) \rho \rho_{23,31} 
    +\Omega_c \rho \rho_{23,21} +\Omega_p (\rho \rho_{13,31} -\rho \rho_{23,32} ) \notag \\
    &-N_a \int d^3 \mathbf{r}^{\prime \prime}\left\langle\hat{\sigma}_{33}\left(\mathbf{r}^{\prime \prime}\right) \hat{\sigma}_{23}\left(\mathbf{r}^{\prime}\right) \hat{\sigma}_{31}(\mathbf{r})\right\rangle V\left(\mathbf{r}^{\prime \prime}-\mathbf{r}\right)=0. \\
    & \left(i \frac{\partial}{\partial t} + d_{32}+ d_{31} -V\left(\mathbf{r}^{\prime}-\mathbf{r}\right) \right) \rho \rho_{32,31} 
    +\Omega_c (\rho \rho_{22,31}+\rho \rho_{32,21}) -\Omega_p (\rho \rho_{31,31} -\rho \rho_{32,32} ) \notag \\
    &-N_a \int d^3 \mathbf{r}^{\prime \prime}\left\langle\hat{\sigma}_{33}\left(\mathbf{r}^{\prime \prime}\right) \hat{\sigma}_{32}\left(\mathbf{r}^{\prime}\right) \hat{\sigma}_{31}(\mathbf{r})\right\rangle V\left(\mathbf{r}^{\prime \prime}-\mathbf{r}\right)=0. \\
    & \left(i \frac{\partial}{\partial t} + d_{21}+ i \Gamma_{23}\right) \rho \rho_{33,21} 
    +\Omega_c \rho \rho_{23,21} +\Omega_p (\rho \rho_{33,11} -\rho \rho_{33,22} )
    +\Omega_c^* \left( \rho \rho_{33,31} -\rho \rho_{32,21} \right)=0. \\
    & \left(i \frac{\partial}{\partial t} + d_{31} + i \Gamma_{12} \right) \rho \rho_{22,31} 
    +\Omega_c (\rho \rho_{22,21} -\rho \rho_{23,31}) +\Omega_p (\rho \rho_{12,31} -\rho \rho_{21,31} -\rho \rho_{22,32}) \notag \\
    &+\Omega_c^* \rho \rho_{32,31} -N_a \int d^3 \mathbf{r}^{\prime \prime}\left\langle\hat{\sigma}_{33}\left(\mathbf{r}^{\prime \prime}\right) \hat{\sigma}_{22}\left(\mathbf{r}^{\prime}\right) \hat{\sigma}_{31}(\mathbf{r})\right\rangle V\left(\mathbf{r}^{\prime \prime}-\mathbf{r}\right)=0. \\
    & \left(i \frac{\partial}{\partial t} + d_{21} + d_{23} \right) \rho \rho_{23,21} 
    +\Omega_p (\rho \rho_{13,21} -\rho \rho_{23,11} -\rho \rho_{23,22}) +\Omega_c^* ( \rho \rho_{23,31} +\rho \rho_{33,21}- \rho \rho_{22,21} ) \notag \\
    &-N_a \int d^3 \mathbf{r}^{\prime \prime}\left\langle\hat{\sigma}_{33}\left(\mathbf{r}^{\prime \prime}\right) \hat{\sigma}_{23}\left(\mathbf{r}^{\prime}\right) \hat{\sigma}_{21}(\mathbf{r})\right\rangle V\left(\mathbf{r}^{\prime \prime}-\mathbf{r}\right)=0. \\
    & \left(i \frac{\partial}{\partial t} + d_{21} + d_{32} \right) \rho \rho_{32,21} +\Omega_c ( \rho \rho_{22,21} - \rho \rho_{33,21} ) +\Omega_p (\rho \rho_{32,11} -\rho \rho_{31,21} -\rho \rho_{32,22}) \notag \\
    &+\Omega_c^* \rho \rho_{32,31} =0. \\
    & \left(i \frac{\partial}{\partial t} + d_{21} + i \Gamma_{12} \right) \rho \rho_{22,21} 
    -\Omega_c \rho \rho_{23,21} +\Omega_p (\rho \rho_{12,21} -\rho \rho_{21,21} -\rho \rho_{22,22} + \rho \rho_{22,11}) \notag \\
    &+\Omega_c^* (\rho \rho_{22,31} + \rho \rho_{32,21}) =0.
    \end{align}
\end{subequations}
\end{widetext}

These equations constitute a closed set of equations for $\rho \rho_{33,31}$, to be more specific we address that one may notice that there are the interaction terms such as $V \left(\mathbf{r}^{\prime}-\mathbf{r}\right) \rho \rho_{32,31}$ is out of the integration. This arises from the delta-function extraction property when commuting projection operators in the integral of the many-body interaction term using commutation relations
\begin{equation}
    [\hat{\sigma}_{\alpha \beta}(\mathbf{r}),\hat{\sigma}_{\mu \nu}(\mathbf{r}^{\prime})]=(\delta_{\alpha \nu}\hat{\sigma}_{\mu \beta}(\mathbf{r})-\delta_{\mu \beta}\hat{\sigma}_{\alpha \nu}(\mathbf{r}^{\prime}))\delta_{\mathbf{rr^{\prime}}}.
\end{equation}

Next, substituting the perturbation expansion and retaining the third-order terms, we obtain the following system of equations $Q\mathbf{x}^{(3)}=\mathbf{q}$:

\begin{widetext}

\begin{equation}
\setlength\arraycolsep{2pt}
\resizebox{\linewidth}{!}{$
    Q= \left[\begin{array}{cccccccc}
    d_{31}+i \Gamma_{23}-V(\mathbf{r^{\prime}}-\mathbf{r}) & \Omega_c & -\Omega_c^* & \Omega_c & 0 & 0 & 0 & 0 \\
    \Omega_c^* & d_{23}+d_{31} & 0 & 0 & -\Omega_c^* & \Omega_c & 0 & 0 \\
    -\Omega_c & 0 & d_{31}+d_{32}-V(\mathbf{r^{\prime}}-\mathbf{r}) & 0 & \Omega_c & 0 & \Omega_c & 0 \\
    \Omega_c^* & 0 & 0 & d_{21}+ i \Gamma_{23} & 0 & \Omega_c & -\Omega_c^* & 0 \\
    0 & -\Omega_c & \Omega_c^* & 0 & d_{31}+i \Gamma_{12} & 0 & 0 & \Omega_c \\
    0 & \Omega_c^* & 0 & \Omega_c^* & 0 & d_{21}+d_{23} & 0 & -\Omega_c^* \\
    0 & 0 & \Omega_c^* & -\Omega_c & 0 & 0 & d_{21}+d_{32} & -\Omega_c^* \\
    0 & 0 & 0 & 0 & \Omega_c^* & -\Omega_c & \Omega_c^* & d_{21}+ i \Gamma_{12} 
\end{array}\right]
  $}
\end{equation}

\begin{equation}
\mathbf{x}^{(3)}= \left[\rho \rho_{33,31}^{(3)} ,\rho \rho_{23,31}^{(3)} ,\rho \rho_{32,31}^{(3)} ,\rho \rho_{33,21}^{(3)} ,\rho \rho_{22,31}^{(3)} ,\rho \rho_{23,21}^{(3)} ,\rho \rho_{32,21}^{(3)} ,\rho \rho_{22,21}^{(3)} \right]^{T},
\end{equation}

\begin{equation}
\mathbf{q} =\left[\begin{array}{c}
0 \\
-\rho \rho_{12,31}^{(2)} \\
\rho \rho_{31,31}^{(2)} \\
-\rho_{33}^{(2)} \\
-\rho \rho_{12,31}^{(2)}+ \rho \rho_{21,31}^{(2)} \\
-\rho_{23}^{(2)}+\rho \rho_{13,21}^{(2)} \\
-\rho_{32}^{(2)}+\rho \rho_{31,21}^{(2)} \\
-\rho_{22}^{(2)}-\rho \rho_{12,21}^{(2)}+\rho \rho_{21,21}^{(2)} 
\end{array}\right],
\end{equation}

\end{widetext}

And for second order terms in $\mathbf{q}$, we follow the previous procedure, write dynamic equations of two-body correlators then retain the second-order terms,
\begin{widetext}
\begin{equation}
\begin{aligned}
& {\left[\begin{array}{cccc}
d_{13}+d_{31} & -\Omega_c^* & 0 & \Omega_c \\
-\Omega_c & d_{12}+d_{31} & \Omega_c & 0 \\
0 & \Omega_c^* & d_{12}+d_{21} & -\Omega_c \\
\Omega_c^* & 0 & -\Omega_c^* & d_{13}+d_{21}
\end{array}\right]\left[\begin{array}{l}
\rho \rho_{13,31}^{(2)} \\
\rho \rho_{12,31}^{(2)} \\
\rho \rho_{12,21}^{(2)} \\
\rho \rho_{13,21}^{(2)} 
\end{array}\right]} \\
& =\left[\begin{array}{c}
0 \\
\rho_{31}^{(1)}\\
-\rho_{12}^{(1)}+\rho_{21}^{(1)} \\
-\rho_{13}^{(1)}
\end{array}\right],
\end{aligned}
\end{equation}

\begin{equation}
\begin{aligned}
& {\left[\begin{array}{cccc}
2d_{31}-V(\mathbf{r}^{\prime}-\mathbf{r}) & 2\Omega_c & 0 & 0 \\
\Omega_c^* & d_{21}+d_{31} & \Omega_c & 0 \\
0 & 0 & 2 d_{21} & 2 \Omega_c^* \\
\Omega_c^* & 0 & \Omega_c & d_{21}+d_{31}
\end{array}\right]\left[\begin{array}{l}
\rho \rho_{31,31}^{(2)} \\
\rho \rho_{21,31}^{(2)} \\
\rho \rho_{21,21}^{(2)} \\
\rho \rho_{31,21}^{(2)} 
\end{array}\right]} \\
& =\left[\begin{array}{c}
0 \\
\rho_{31}^{(1)} \\
-2 \rho_{21}^{(1)} \\
-\rho_{31}^{(1)}
\end{array}\right],
\end{aligned}
\end{equation}

\end{widetext}

\section*{Acknowledgments}
We thank Yong-Chang Zhang for helpful discussions. Jia-Wei Lai acknowledges the support of  National Natural Science Foundation of China (Grant Nos. 12404389), the Natural Science Basic Research Program of Shaanxi (Program No. 2024JC-YBQN-0063). Pei Zhang acknowledges the support of the National Nature Science Foundation of China (Grant No. 12174301), the Natural Science Basic Research Program of Shaanxi (Grant No. 2023-JC-JQ-01).\\

\bibliographystyle{apsrev4-2}
\bibliography{biblio}

\end{document}